\documentclass[conference]{IEEEtran}
\IEEEoverridecommandlockouts
\usepackage{cite}
\usepackage{amsmath,amssymb,amsfonts}
\usepackage{algorithmic}
\usepackage{graphicx}
\usepackage{textcomp}
\usepackage{xcolor}
\usepackage{bbm}
\usepackage{fancyhdr}
\usepackage{hyperref}
\def\BibTeX{{\rm B\kern-.05em{\sc i\kern-.025em b}\kern-.08em
    T\kern-.1667em\lower.7ex\hbox{E}\kern-.125emX}}

\usepackage{float}
\usepackage[utf8]{inputenc}
\usepackage[english]{babel}
\usepackage{lettrine}
\newtheorem{axiom}{Axiom}
\newtheorem{definition}{Definition}

\newtheorem{remark}{Remark}

\newcommand{\indep}{\perp \!\!\! \perp}

\makeatletter
 \let\old@ps@headings\ps@headings
 \let\old@ps@IEEEtitlepagestyle\ps@IEEEtitlepagestyle
 \def\confheader#1{%
 \def\ps@IEEEtitlepagestyle{%
 \old@ps@IEEEtitlepagestyle%
 \def\@oddhead{\strut\hfill#1\hfill\strut}%
 \def\@evenhead{\strut\hfill#1\hfill\strut}%
 }%
 \ps@headings%
 }
 \makeatother

\confheader{%
INTERNATIONAL JOURNAL OF CIRCUITS, SYSTEMS AND SIGNAL PROCESSING, Volume 15, 2021
 }

\begin{document}

\title{Decision Trees with Soft Numbers \\
}

\author{\IEEEauthorblockN{Oren Fivel}
\IEEEauthorblockA{\textit{School of Electrical and Computer Engineering} \\
\textit{Ben-Gurion University of the Negev}\\
Be'er Sheva, Israel \\
fivel@post.bgu.ac.il}
\and
\IEEEauthorblockN{Moshe Klein}
\IEEEauthorblockA{\textit{Dep. of Industrial Engineering} \\
\textit{Tel-Aviv University}\\
Tel-Aviv, Israel \\
mosheklein@mail.tau.ac.il}
\and
\IEEEauthorblockN{Oded Maimon}
\IEEEauthorblockA{\textit{Dep. of Industrial Engineering} \\
\textit{Tel-Aviv University}\\
Tel-Aviv, Israel \\
maimon@tauex.tau.ac.il}
}

\twocolumn[
  \begin{@twocolumnfalse}
      \date{}
     \maketitle

\hspace {1cm}\begin{minipage}{0.9\linewidth}
Received: July 16, 2021.  Revised: November 25, 2021. Accepted: December 15, 2021.  Published: January  2, 2022.\\
DOI: 
\href{https://www.naun.org/main/NAUN/circuitssystemssignal/2021/d942005-194(2021).php}{10.46300/9106.2021.15.194}\\
\end{minipage}

     \end{@twocolumnfalse}
]


\begin{abstract}
In this paper we develop the foundation of a new theory for decision trees based on new modeling of phenomena with soft numbers. Soft numbers represent the theory of soft logic that addresses the need to combine real processes and cognitive ones in the same framework. At the same time soft logic develops a new concept of modeling and dealing with uncertainty: the uncertainty of time and space. It is a language that can talk in two reference frames, and also suggest a way to combine them.
In the classical probability, in continuous random variables there is no distinguishing between the probability involving strict inequality and non-strict inequality. Moreover, a probability involves equality collapse to zero, without distinguishing among the values that we would like that the random variable will have for comparison. This work presents \textit{Soft Probability}, by incorporating of \textit{Soft Numbers} into probability theory. \textit{Soft Numbers} are set of new numbers that are linear combinations of multiples of "ones" and multiples of "zeros". In this work, we develop a probability involving equality as a "soft zero" multiple of a probability density function (PDF). We also extend this notion of soft probabilities to the classical definitions of Complements, Unions, Intersections and Conditional probabilities, and also to the expectation, variance and entropy of a continuous random variable, condition being in a union of disjoint intervals and a discrete set of numbers. This extension provides information regarding to a continuous random variable being within discrete set of numbers, such that its probability does not collapse completely to zero. When we developed the notion of soft entropy, we found potentially another soft axis, multiples of 0log(0), that motivates to explore the properties of those new numbers and applications. We extend the notion of soft entropy into the definition of Cross Entropy and Kullback–Leibler-Divergence (KLD), and we found that a soft KLD is a soft number, that does not have a multiple of 0log(0). Based on a soft KLD, we defined a soft mutual information, that can be used as a splitting criteria in decision trees with data set of continuous random variables, consist of single samples and intervals.\\ 
\end{abstract}


\section{Introduction}
\lettrine{\bfseries I}{N} this paper we develop the foundation of a new theory for decision trees based on new modeling of phenomena with soft numbers. This calls for major concept change of probability, which is developed in this paper, so that decision trees can be modeled. Soft numbers represent the theory of soft logic that addresses the need to combine real processes and cognitive ones in the same framework. At the same time soft logic develops a new concept of modeling and dealing with uncertainty: the uncertainty of time and space. It is a language that can talk in two reference frames, and also suggest a way to combine them.
\subsection{Research Motivation and Direction}
Probability theory is used in order to model processes and phenomenons, involving randomness of the parameters and variables (See Appendix \ref{Appendix A} for a brief review and notations regarding to probability theory). A probability a continues random variable is defined by a Probability Density Function (PDF). The PDF can be used is to approximate the probability of the continuous random variable $X$ to be adjacent to $x$ in the following sense
\begin{equation}\label{Prob_approx}
\mathrm{Pr}(x<X\leq x+\Delta x) \approx f_X(x)\Delta x,  
\end{equation}
where $\Delta x>0$ is  a small value, that defines how much this probability is accurate. However, continuous random variables have the following properties:
\begin{itemize}
\item No distinguishing between strict inequality and non-strict in equality e.g., $\mathrm{Pr}(X\leq x)=\mathrm{Pr}(X<x)$;
\item Equality collapses to zero i.e.,  $\mathrm{Pr}(X=x)=0$. Although any value of $x\in S_X$ ($S_X$ denotes the support of $X$) is possible for $X$, the the probability of $X$ to be equal to any value of $x\in S_X$ is (almost surely) zero.  
\end{itemize}
Because of these properties, we lose some information regarding to a continuous random variable to have an exact value.
On one hand, an event "$X=x$" might be possible (if $x\in S_X$) but improbable (i.e., with zero probability), which seems to be a paradox. On the other hand, we can express the zero probability by of an event "$X=x$" by letting $\Delta x$ to approach to zero in \eqref{Prob_approx}
\begin{equation}\label{Prob_zero_event}
\mathrm{Pr}(X=x) = f_X(x)\cdot 0.  
\end{equation}
This equation presents the probability $\mathrm{Pr}(X=x)$ as a multiple of zero with a factor of the PDF $f_X(x)$ for all $x$. Instead of taking $\mathrm{Pr}(X=x)$ to be completely zero, we can assign to it a zero multiple of $f_X(x)$ and compare different probability values for different observation values $x$. This approach can be implemented by using \textit{Soft Numbers} (see Appendix \ref{Appendix B} and Klein and Maimon's papers e.g., \cite{Klein_Maimon1}, \cite{Klein_Maimon2} and \cite{Klein_Maimon3}). 

In addition there is an approach to represent a discrete distribution as a continuous distribution by a linear combination of Dirac delta functions $\delta(x-x_i)$, or by any approximations of Dirac delta functions e.g., Gaussian functions (also known as Gaussian mixture model or GMM) or rectangular functions (based on uniform distribution) etc (see \eqref{discrete_t_delta} for more details). Our approach it to establish the opposite in some sense, i.e., to represent a continuous random variable with a possibility to have a discrete values with probability that will not collapse absolutely to zero.

In this work, we introduce the \textit{Soft Numbers} to give a probability interpretation of a continuous random variable to have an exact value, that provides distinguishing between strict inequality and non-strict in equality in the probability function.

\subsection{Organization of the Work}
Section \ref{sec2} incorporates Soft Numbers into probability theory to present the notion of "Soft Probability". Section \ref{sec3} extends this notion to conditional probability. Section \ref{sec4} defines a Soft Expectation, a Variance and a Soft Entropy, where the last generates potentially another soft axis, multiples of $0\cdot\log0$. Section \ref{sec5} presents an example for application on Decision Trees based on a Soft Mutual Information as a Splitting Criteria. Conclusions and suggestion for future research are shown on sections \ref{sec6} and \ref{sec7} respectively to summarize this work. For completion, Appendix \ref{Appendix A} provides a brief review of probability theory, and Appendix \ref{Appendix B} provides a presentation of Soft Numbers.

\section{Soft Probability: Incorporation of Soft Number into Probability Theory}\label{sec2}
In order to incorporate the notion of  \eqref{softNumFunc} in Appendix \ref{Appendix B}, we define  \eqref{CDF} in Appendix \ref{Appendix A} differently for a cumulative distribution function (CDF) of a continuous random variable
\begin{equation}\label{Soft_P=CDF}
\mathrm{Ps}(X\leq x)=F_X(1\cdot \bar{0}\dot{+}x), 
\end{equation}
where $\mathrm{Ps}(\cdot)$ is a suggested type of a probability function, dented as a "Soft Probability" [instead of a regular probability notation "$\mathrm{Pr}(\cdot)$" or $P(\cdot)$], and $F_X(\cdot)$ is the regular CDF function of the random variable $X$ but it is applied on a soft number $1\cdot \bar{0}\dot{+}x$. Our motivation is to generate an alternative evaluation of the probability at the left hand side (LHS), so that we can distinguish between $\mathrm{Ps}(X<x)$ and $\mathrm{Ps}(X\leq x)$ for a continuous random variable $X$ [i.e., $\mathrm{Ps}(X<x) \neq \mathrm{Ps}(X\leq x)$]. We will show that the evaluation of the soft number at the CDF in the right hand side (RHS) will create this distinction.

The RHS of \eqref{Soft_P=CDF} can be decomposed by \eqref{softNumFunc} as follows
\begin{equation}\label{soft_CDF}
F_X(1\cdot \bar{0} \dot{+}x)\overset{\textrm{def}}{=} f_X(x)\bar{0} \dot{+}F_X(x),
\end{equation}
The LHS of \eqref{Soft_P=CDF} can be decomposed by separating the event $"X\leq x"$ into a disjoint union $"X=x\uplus X<x"$. In a regular probability, we have the known identities
\begin{align*}
 \mathrm{Pr}(X\leq x)&\overset{"X=x"\cap "X<x"=\emptyset}{=}
 \underbrace{\mathrm{Pr}(X=x)}_{=0}+\mathrm{Pr}(X<x)\\
 &=\mathrm{Pr}(X<x),
\end{align*}
So we do not have a distinction between $\mathrm{Pr}(X\leq x)$ and $\mathrm{Pr}(X < x)$.
We distinguish between $\mathrm{Ps}(X\leq x)$ and $\mathrm{Ps}(X < x)$ by the following definition for $\mathrm{Ps}(X\leq x)$
\begin{equation}\label{Soft_P}
\mathrm{Ps}(X\leq x)\overset{\textrm{def}}{=}\mathrm{Ps}(X=x)+\mathrm{Ps}(X<x), 
\end{equation}
so that we  define the terms on the LHS as follows

\begin{equation}\label{Soft_P=af}
\mathrm{Ps}(X=x)\overset{\textrm{def}}{=}f_X(x)\bar{0}, \\
\end{equation}

\begin{equation}\label{Soft_P<af}
\mathrm{Ps}(X<x)\overset{\textrm{def}}{=}F_X(x) \equiv \mathrm{Pr}(X<x).
\end{equation}
By this setup we achieve a distinguishing between $\mathrm{Ps}(X\leq x)$ and $\mathrm{Ps}(X<x)$, an also we provide an interpretation to $\mathrm{Ps}(X=x)$ be infinitesimally small but not collapse completely to zero due to the factor $\bar{0}$ of the PDF.

In the next subsection, we provide two examples of implementations on PDFs, Gaussian distribution and uniform distribution, in order to demonstrate the effect of soft numbers (and more precisely, soft zeros) on PDFs.     

\subsection{Examples}
\subsubsection{Gaussian distribution}

Let $X$ be a Gaussian random variable parameterized by a mean $\mu$ and a variance $\sigma ^2$ [denoted $X\sim N(\mu,\sigma ^2)$]. The PDF of $X$ is well known as
\begin{equation}
f_X(x;\mu , \sigma ^2)=\frac{1}{\sqrt{2\pi \sigma ^2}}e^{-\frac{1}{2\sigma ^2}(x-\mu )^2}.
\end{equation}

The maximum of the PDF, $\max_x f_X(x;\mu ,\sigma ^2)=\frac{1}{\sqrt{2\pi \sigma ^2}}$, occurs at $x=\mu$. We would like to have a high probability as $X$ is closer to $\mu $ e.g., $\mathrm{Ps}(X=\mu)>\mathrm{Ps}(X=x), \forall x\neq \mu$. 
By  \eqref{Soft_P=af} and we have the following definition a soft probability in the Gaussian case

\begin{equation}
\mathrm{Ps}(X=x;\mu , \sigma ^2)= \frac{1}{\sqrt {2\pi \sigma ^2}} e^{-\frac{1}{2\sigma ^2}(x-\mu )^2} \cdot\bar{0},
\end{equation}
which presents an absolute low probability of $X$ to have an exact value $x$ but relative high probability when $X$ is closer to $\mu$. 

\subsubsection{Uniform distribution}
Let $X$ be Uniformly distributed at the interval $(a,b)$  [denoted $X\sim U(a,b)$]. The PDF of $X$ is well known as
\begin{equation}
f_X(x;a , b)=\frac{1}{b-a}\mathbbm{1}_{x\in (a,b)},
\end{equation}
where $\mathbbm{1}_A$ is the indication function, indicates '1' if '$A$' is true and '0' if '$A$' is false. Similarly to previous example (but with maximal PDF to be trivially $\frac{1}{b-a}$) we have the following soft probability in the uniform case
\begin{equation}
\mathrm{Ps}(X=x;a , b)=\frac{1}{b-a}\cdot\mathbbm{1}_{x\in (a,b)}\cdot\bar{0},
\end{equation}
which implies the following property:
\begin{equation}
\begin{split}
&\forall x,y\in\mathbb{R}, x\in(a,b) \wedge y\notin(a,b) \\ 
&\Rightarrow \mathrm{Ps}(X=x)=\frac{1}{b-a}\cdot\bar{0}>\mathrm{Ps}(X=y)=0\cdot\bar{0}. 
\end{split}
\end{equation}
This property emphasises the probability to $X$ to have any value within $(a,b)$ is absolutely small, but still relative greater than the probability to have any value outside of which is almost surely impossible). 

\subsection{Observations}
In soft numbers development, we may consider to distinct between two options to define an absolute value of a soft number: Option 1 is by the definition in  \eqref{softNumFunc} with $|x|'=\textrm{sign}(x)$, ignoring the fact that this derivative is not continuous, so that
\begin{equation}\label{soft abs option 1}
    |a\bar{0} \dot{+}x|=\alpha \bar{0} \cdot \textrm{sign} (x)\dot{+}|x|.
\end{equation}
Option 2 is to define a soft conjugate of $\alpha \bar{0} \dot{+}x$ to be  $(-\alpha)\bar{0} \dot{+}x$ such that
\begin{equation}\label{soft abs option 2}
\begin{split}
|\alpha \bar{0} \dot{+}x| &=\sqrt{(\alpha \bar{0} \dot{+}x)((-\alpha)\bar{0} \dot{+}x)}\\
&=\sqrt{-(\alpha \bar{0})^2 + x^2}\\
&=\sqrt{-0 + x^2}\\
&=\sqrt{x^2}\\
&=|x|.
\end{split}
\end{equation}
If we use Option 2, then we can have the following properties for a soft probability on a continuous random variable:
\begin{enumerate}
\item $\mathrm{Ps}(X\leq x)\neq \mathrm{Ps}(X< x)$\\
but $|\mathrm{Ps}(X\leq x)|=|\mathrm{Ps}(X< x)|>|\mathrm{Ps}(X= x)|=0 $;
\item $f_X(x)>f_X(y) \Rightarrow \mathrm{Ps}(X= x)>\mathrm{Ps}(X= y)$\\
but $|\mathrm{Ps}(X= x)|=|\mathrm{Ps}(X= y)|=0$;
\item $f_X(x)>f_Y(y) \Rightarrow \mathrm{Ps}(X= x)>\mathrm{Ps}(Y= y)$\\
but $|\mathrm{Ps}(X= x)|=|\mathrm{Ps}(Y= y)|=0$;
\item $|\mathrm{Ps}(X\leq x)|=\mathrm{Ps}(X< x)=\mathrm{Pr}(X< x)=\\\mathrm{Pr}(X\leq x)$.
\end{enumerate}
By taking absolute values of the soft probability term, we return to the classic probability results for continuous random variable e.g., not distinguishing between strict inequality and non-strict inequality, and equality collapse to zero.

In the next section, we extend the notion of "Soft Probability" into the events' complements, unions and intersections , and into conditional probability.

\section{Complements, Union, Intersection and Conditional Soft Probability}\label{sec3}
In the following section we extend the notion of "Soft Probability" into the events' complements, unions and intersections, and into conditional probability. In the First Subsection we show the results for complements, unions and intersections corresponding to event with zero probability in the classical probability sense. In the second subsection we show the results for a conditional of soft probability, referring to Kolmogorov definition and Bayes theorem.    

\subsection{Complements, Unions and Intersections}
Recall that a probability of $A^c$, a complement of the event $A$, is given by
\begin{equation}
\mathrm{Pr}(A^c)=1-\mathrm{Pr}(A).
\end{equation}
A Soft probability of a complement is defined similarly as follows
\begin{equation}
\mathrm{Ps}(A^c)=1-\mathrm{Ps}(A).
\end{equation}
Therefore, we have the following probability complement for a continuous random variable $X$:
\begin{equation}
\begin{split}
\mathrm{Ps}(X\neq x)&=1-\mathrm{Ps}(X=x)\\&=[-f_X(x)]\bar{0}\dot{+}1.
\end{split}
\end{equation}
This equation distinguishes among different values of $x$ for the event $X\neq x$ to be with almost surely with probability 1 due the the soft zero term $[-f_X(x)]\bar{0}$. This equation is analogous to the event $X\neq x$ to have zero probability almost surely, correct by the soft zero term $[-f_X(x)]\bar{0}$.

In order to analyse unions and intersections, we need to consider two cases: unions and intersections among singletons events $X=x, X=y$ etc; unions and intersections between a singleton event $X=x$ and a range event e.g. $a \leq X \leq b$.

For all $x\neq y$ we have that the events $X=x$ and $X=y$ are disjoint, and the for a union we have
\begin{equation}
\begin{split}
\mathrm{Ps}(X=x\cup X=y)&=\mathrm{Ps}(X=x)+\mathrm{Ps}(X=y)\\
              &=[f_X(x)+f_X(y)]\bar{0}.
\end{split}
\end{equation}
For an intersection we have
\begin{equation}
\mathrm{Ps}(X=x\cap X=y)=\mathbbm{1}_{x=y}f_X(x)\bar{0},
\end{equation}
where the indicator $\mathbbm{1}_{x=y}$ is zero in the case that $x\neq y$. More generally, we have the following soft probabilities for the following set $\{x_i\}_{i=1}^n$ with distinct values:

\begin{equation}\label{soft n equal union}
\mathrm{Ps}\left (\bigcup_{i=1}^{n}X=x_i\right )= \sum_{i=1}^{n}\mathrm{Ps}(X=x_i)=\left [ \sum_{i=1}^{n}f_X(x_i) \right ]\bar{0},
\end{equation}
and 
\begin{equation}\label{soft n equal intersect}
\mathrm{Ps}\left (\bigcap_{i=1}^{n}X=x_i\right )
=\mathbbm{1}_{x_i=x_j}^{\forall i,j\in \left \{ 1,2,...,n \right \}}f_X(x_i)\bar{0}.
\end{equation}

In order to analyse unions and intersections, between a singleton event $X=x$ and a range event e.g. $a \leq X \leq b$, we need to distinguish among $x$'s values that are either between $a$ and $b$ or not. Moreover we need to distinguish between the strict inequality case $a < X < b$ and the non-strict inequality $a \leq X \leq b$. For simplicity, assume $a<b$ and without loss of generality (WLOG) assume $x\neq a$ and $x\neq b$. 

For the strict inequality case $a < X < b$ we have the union 
\begin{equation}\label{strict union}
\mathrm{Ps}(X=x\cup a<X<b)=\mathbbm{1}_{x\notin (a,b)}f_X(x)\bar{0} \dot{+}  [F_X(b)-F_X(a)],
\end{equation}
and for the intersection
\begin{equation}\label{strict intesect}
\mathrm{Ps}(X=x\cap a<X<b)=\mathbbm{1}_{x\in (a,b)}f_X(x)\bar{0}.
\end{equation}
This union is a soft number when $x$ is not in the interval $(a,b)$ and a real number when it does. This intersection is a soft zero when $x$ is in $(a,b)$ and an absolute zero when it doesn't.

For the non-strict inequality case $a \leq X \leq b$ we have the union 
\begin{equation}\label{non-strict union}
\begin{split}
&\mathrm{Ps}(X=x\cup a \leq X \leq b)=\\
& [\mathbbm{1}_{x\notin [a,b]}f_X(x)+f_X(a)+f_X(b)]\bar{0} \dot{+}  [F_X(b)-F_X(a)],
\end{split}
\end{equation}
and for the intersection
\begin{equation}\label{non-strict intersect}
\mathrm{Ps}(X=x\cap a \leq X \leq b)=[\mathbbm{1}_{x\in [a,b]}f_X(x)]\bar{0}.
\end{equation}
the two terms $f_X(a)+f_X(b)$ in  \eqref{non-strict union} are added to the soft zero part, due to  \eqref{soft n equal union}. 

Recall the relation between a union and an intersection of two events $A,B$, according to De Morgan's Law, we have
\begin{equation}\label{De Morgan's Prob}
\mathrm{Pr}(A\cup B)=\mathrm{Pr}(A)+\mathrm{Pr}(B)-\mathrm{Pr}(A\cap B).
\end{equation}
It can be shown that the soft probabilities in \eqref{strict union}-\eqref{non-strict intersect} hold for De Morgan's Law  \eqref{De Morgan's Prob}. For example $A=\left \{ X=x \right \}$, $B=\left \{ a\leq X\leq b \right \}$ and $x\notin [a,b]$, we have

\begin{equation}\label{Soft De Morgan's Prob X}
\begin{split}
&\mathrm{Ps}(X=x\cup a\leq X\leq b)=\\
&\mathrm{Ps}(X=x)+\mathrm{Pr}(a\leq X\leq b)-\mathrm{Pr}(X=x\cap a\leq X\leq b).
\end{split}
\end{equation}
The LHS is
\[[f_X(x)+f_X(a)+f_X(b)]\bar{0} \dot{+}  [F_X(b)-F_X(a)]\]
and the RHS is
\[f_X(x)\bar{0}+\left [ \left \{ f_X(a)+f_X(b) \right \} \bar{0} \dot{+} \{F_X(b)-F_X(a)\} \right ]-0,\]
so that we obtain the LHS to be equal to the RHS, and thus we have a "Soft De Morgan's Law"
\begin{equation}\label{Soft De Morgan's Prob}
\mathrm{Ps}(A\cup B)=\mathrm{Ps}(A)+\mathrm{Ps}(B)-\mathrm{Ps}(A\cap B).
\end{equation}

In the next subsection, we show the results for a conditional of soft probability, referring to Kolmogorov definition and Bayes theorem. 
\subsection{Conditional Probability}
Recall Kolmogorov definition for conditional probability 
\begin{equation}\label{Kolmogorov Condition Prob}
\mathrm{Pr}(A|B)=\frac{\mathrm{Pr}(A\cap B)}{\mathrm{Pr}(B)}, 
\end{equation}
and for Bayes theorem
\begin{equation}\label{Bayes theorem}
\mathrm{Pr}(A|B)=\frac{\mathrm{Pr}(B|A)\mathrm{Pr}(A)}{\mathrm{Pr}(B)}, 
\end{equation}

We define a "Soft Conditional Probability" similarly, e.g., for $x,y\in S_X$, let $A=\left \{ X=x \right \}$, $B=\left \{ X=y \right \}$, and at the LHS of Kolmogorov definition \eqref{Kolmogorov Condition Prob} we have
\begin{equation}
\frac{\mathrm{Ps}(X=x\cap X=y)}{\mathrm{Ps}(X=y)}=\frac{\mathbbm{1}_{x=y}f_X(x)\bar{0}}{f_X(y)\bar{0}}=\frac{\mathbbm{1}_{x=y}\cdot \bar{0}}{1 \cdot \bar{0}}.
\end{equation}

With a definition of $\frac{1\cdot \bar{0}}{1 \cdot \bar{0}}=1$ and $\frac{0\cdot \bar{0}}{1 \cdot \bar{0}}=0$, the conditional soft probability is given by 
\begin{equation}\label{Kolmogorov Soft Equality}
\mathrm{Ps}(X=x|X=y)=\mathbbm{1}_{x=y}. 
\end{equation}
In this case we have a trivial equality with optional real values 0 or 1. For comparison with Bayes theorem \eqref{Bayes theorem}
\begin{equation}
\frac{\mathrm{Ps}(X=y|X=x)\mathrm{Ps}(X=x)}{\mathrm{Ps}(X=y)}=\frac{\mathbbm{1}_{y=x}f_X(x)\bar{0}}{f_X(y)\bar{0}}=\mathbbm{1}_{x=y}.
\end{equation}
Now we consider $x,y\in S_X$, let $A=\left \{ X=x \right \}$, $B=\left \{ a\leq X\leq b \right \}$, with $x,a,b\in S_X$ such that $a<b$ ,$x\neq a$ and $x\neq b$.
At the LHS of Kolmogorov definition \eqref{Kolmogorov Condition Prob} we have

\begin{equation}
\begin{split}
&\frac{\mathrm{Ps}(X=x\cap a\leq X\leq b)}{\mathrm{Ps}(a\leq X\leq b)}=
\\
&\frac{\mathbbm{1}_{x\in [a,b]}f_X(x)\bar{0}}{[f_X(a)+f_X(b)]\bar{0} \dot{+} [F_X(b)-F_X(a)]}.
\end{split}
\end{equation}
When applying Bayes theorem \eqref{Bayes theorem}, we have
\begin{equation}
\begin{split}
&\frac{\mathrm{Ps}(a\leq X\leq b|X=x)\mathrm{Ps}(X=x)}{\mathrm{Ps}(a\leq X\leq b)}=\\
&\frac{[\mathrm{Ps}(a\leq x\leq b|X=x)]f_X(x)\bar{0}}{[f_X(a)+f_X(b)]\bar{0} \dot{+} [F_X(b)-F_X(a)]},
\end{split}
\end{equation}
where $\mathrm{Ps}(a\leq x\leq b|X=x)=\mathrm{Ps}(a\leq x\leq b)=\mathbbm{1}_{x\in [a,b]}$. Both Kolmogorov theorem form and Bayes theorem form are equal, and therefore 
\begin{equation}
\mathrm{Ps}(X=x|a\leq X\leq b)
=\frac{\mathbbm{1}_{x\in [a,b]}f_X(x)\bar{0}}{[f_X(a)+f_X(b)]\bar{0} \dot{+} [F_X(b)-F_X(a)]}.
\end{equation}
We can simplify the RHS by the property
\[ \frac{A\bar{0}}{B\dot{+}C\bar{0}}= \\ \frac{A\bar{0}}{B\dot{+}C\bar{0}}\cdot \frac{B\dot{+}(-C)\bar{0}}{B\dot{+}(-C)\bar{0}}=\frac{AB\bar{0}}{B^2}=\frac{A\bar{0}}{B},\]
and we have the following conditional soft probability with a given non-strict inequality condition:
\begin{equation}\label{Cond SoftProv nonstrctineq}
\mathrm{Ps}(X=x|a\leq X\leq b)
=\frac{\mathbbm{1}_{x\in [a,b]}f_X(x)\bar{0}}{F_X(b)-F_X(a)},
\end{equation}
and for a given strict inequality condition, we have.
\begin{equation}\label{Cond SoftProv strctineq}
\mathrm{Ps}(X=x|a<X<b)
=\frac{\mathbbm{1}_{x\in (a,b)}f_X(x)\bar{0}}{F_X(b)-F_X(a)}.
\end{equation}
The meaning of these last two equation is that we have a soft zero when the observation $x$ makes sense (i.e. between $a$ and $b$), and it is an absolute zero if $x$ makes no sense (i.e. not between $a$ and $b$), due to the indicator in the numerator. In addition, division by the denominator $F_X(b)-F_X(a)\in (0,1)$ makes higher probability than the unconditional probability, which make sense since we have an additional information regarding to the random variable $X$ to be between $a$ and $b$. 
In the next subsection, we extend the notion of soft probability for 2 continuous random variables, based on a Soft De Morgan's Law \eqref{Soft De Morgan's Prob}. 

\subsection{Extension of Soft Probability for 2 Dimensions}
Suppose that $X$ and $Y$ are two continuous random variables. By the regular De Morgan's Law \eqref{De Morgan's Prob}, we can decompose the regular probability object $\mathrm{Pr}(X\leq x, Y\leq y)$ into a sum of the following probabilities
\begin{equation}
\begin{split}
&\mathrm{Pr}(X\leq x, Y\leq y)=\\
&\overbrace{[\mathrm{Pr}(X<x, Y=y)+\mathrm{Pr}(X=x, Y<y)+\mathrm{Pr}(X=x, Y=y)]}^{0}\\
&+\mathrm{Pr}(X< x, Y< y),
\end{split}    
\end{equation}
such that each of the first three terms in the bracket collapses to zero in the classical probability. We define the soft probability object $\mathrm{Ps}(X\leq x, Y\leq y)$ in 2 random variables based on a Soft De Morgan's Law \eqref{Soft De Morgan's Prob} as follows 
\begin{equation}\label{soft prob 2 dim}
\begin{split}
&\mathrm{Ps}(X\leq x, Y\leq y)=\\
&[\mathrm{Ps}(X<x, Y=y)+\mathrm{Ps}(X=x, Y<y)+\mathrm{Ps}(X=x, Y=y)]\\&+\mathrm{Ps}(X< x, Y< y).
\end{split}    
\end{equation}
In this case, we define the first three terms in the bracket as the following soft zero objects in terms of the CDF $F_{X,Y}(x,y)$ and the PDF $f_{X,Y}(x,y)$:

\begin{equation}\label{Ps(X<x, Y=y)}
\mathrm{Ps}(X<x, Y=y)=\frac{\partial{F_{X,Y}(x,y)}}{\partial y}\cdot\bar{0},
\end{equation}

\begin{equation}\label{Ps(X=x, Y<y)}
\mathrm{Ps}(X=x, Y<y)=\frac{\partial{F_{X,Y}(x,y)}}{\partial x}\cdot\bar{0},
\end{equation}

\begin{equation}\label{Ps(X=x, Y=y)}
\mathrm{Ps}(X=x, Y=y)=\frac{\partial{F_{X,Y}(x,y)}}{\partial x \partial y}\cdot\bar{0}=f_{X,Y}(x,y)\cdot\bar{0}.
\end{equation}
the last term is a regular probability along the 1-axis i.e., 
\begin{equation}\label{Ps(X<x, Y<y)}
\mathrm{Ps}(X<x, Y<y)=\mathrm{Pr}(X<x, Y<y)=F_{X,Y}(x,y),
\end{equation}
so that $\mathrm{Ps}(X\leq x, Y\leq y)$ equals to the following soft number 
\begin{equation}\label{Ps(X<=x, Y<=y}
\begin{split}
&\mathrm{Ps}(X\leq x, Y\leq y)=\\
&\left[\frac{\partial{F_{X,Y}(x,y)}}{\partial x}+\frac{\partial{F_{X,Y}(x,y)}}{\partial y}+f_{X,Y}(x,y)\right]\cdot\bar{0}\\&\dot{+}F_{X,Y}(x,y).
\end{split}    
\end{equation}

Now, we want to construct the soft probability objects $\mathrm{Ps}(X\leq x, Y<y)$ and $\mathrm{Ps}(X\leq x, Y=y)$ [by symmetry, we can construct $\mathrm{Ps}(X< x, Y\leq y)$ and $\mathrm{Ps}(X=x, Y\leq y)$ accordingly].
Based on a Soft De Morgan's Law \eqref{Soft De Morgan's Prob}, we construct the soft probability $\mathrm{Ps}(X\leq x, Y<y)$ similarly as follows:
\begin{equation}\label{Ps(X<=x, Y<y}
\mathrm{Ps}(X\leq x, Y< y)=
\frac{\partial{F_{X,Y}(x,y)}}{\partial x}\cdot\bar{0} \dot{+}F_{X,Y}(x,y).
\end{equation}
Therefore, we can distinguish among the soft probabilities: $\mathrm{Ps}(X\leq x, Y\leq y)$, $\mathrm{Ps}(X<x, Y<y)$, $\mathrm{Ps}(X\leq x, Y<y)$ and $\mathrm{Ps}(X<x, Y\leq y)$. Similarly, we have
\begin{equation}\label{Ps(X<=x, Y=y}
\mathrm{Ps}(X\leq x, Y= y)=
\left[\frac{\partial{F_{X,Y}(x,y)}}{\partial y}+f_{X,Y}(x,y)\right]\cdot\bar{0},
\end{equation}
that is a soft zero.
In the next section, we define soft expectation, soft variance and soft entropy.

\section{Soft Expectation, Variance and Entropy}\label{sec4}
In this section, we define soft expectation, soft variance and soft entropy. First, we focus on expectation and variance's definitions, recalling their original and known definition and then generalizing then to soft numbers. Second, we do this original definition's recall and soft numbers' generalization to the entropy.

\subsection{Soft Expectation and Variance}
Recall for the definition of the Expectation of a random variable $X$ with support $S_X$
\begin{equation}\label{Expectation_def}
\mathrm{E}(X)
=\int_{S_X}xdF_x(x)=\mu_X,
\end{equation}
where $\mathrm{E}(\cdot)$ is the expectation operator defined by the \textit{Lebesgue integral} above, and we denote its result by $\mu_X$ (sometime we call it \textit{mean}). For a continuous random variable the Expectation is defined by
\begin{equation}\label{Expectation_CRV_def}
\mathrm{E}(X)
=\int_{S_X}xf_X(x)dx,
\end{equation}
and for a discrete random variable
\begin{equation}\label{Expectation_DRV_def}
\mathrm{E}(X)
=\sum_{x\in S_X}x\mathrm{Pr}(X=x).
\end{equation}
The Variance of a random variable is the expectation of the square error from its mean, that is
\begin{equation}\label{Variance_def}
\textrm{Var}(X)=\mathrm{E}[(X-\mu)^2]
=\int_{S_X}(x-\mu_X)^2dF_x(x)=\sigma^2_X,
\end{equation}
\begin{equation}\label{Variance_CRV_def}
\textrm{Var}(X)
=\int_{S_X}(x-\mu_X)^2f_X(x)dx, \textrm{for continuous case},
\end{equation}
\begin{equation}\label{Variance_DRV_def}
\textrm{Var}(X)
=\sum_{x\in S_X}(x-\mu_X)^2 \mathrm{Pr}(X=x), \textrm{for discrete case}.
\end{equation}

Suppose that $X$ is a continuous random variable and $\{x_i\}_{i=1}^n$ and $\{(a_j,b_j)\}_{j=1}^m$ are set of numbers and set of disjoint intervals in the support $S_X$. Assume also that $\{x_i\}_{i=1}^n$ and $\{(a_j,b_j)\}_{j=1}^m$ are disjoint. WLOG we consider open intervals, otherwise we can exclude the end point $a_j, b_j$ from the interval $(a_j,b_j)$ and include then into the set of numbers $\{x_i\}_{i=1}^n$. Under the above assumption we define a soft expectation of $X$ as the expectation of $X$ conditioned by being within the union of $\{x_i\}_{i=1}^n$ and $\{(a_j,b_j)\}_{j=1}^m$, i.e.,
\begin{equation}\label{soft_Expectation_def}
\begin{split}
&\mathrm{Es}(X|X\in \{x_i\}_{i=1}^n \cup \{(a_j,b_j)\}_{j=1}^m )\\ &=\sum_{i=1}^nx_i\mathrm{Ps}(X=x_i) \dot{+} \sum_{j=1}^m \int_{a_j}^{b_j} xf_X(x)dx\\ 
 &= \sum_{i=1}^nx_if_X(x_i)\cdot\bar{0} \dot{+} \sum_{j=1}^m\int_{a_j}^{b_j}xf_X(x)dx\\
 &=\nu_X\bar{0}  \dot{+} \kappa_X,
\end{split}
\end{equation}
where $\mathrm{Es}(\cdot)$ is a new notation for a Soft Expectation operator. Here we use the concept of a Conditional Expectation, however instead of calculating an expectation of a random variable given another random variable (see e.g., \cite{{Lando_Ortobelli}}), the condition is given on the same variable but being within some set of single point and intervals. Recall that $X$ is a random variable with a real value, and also all the single point $\{x_i\}_{i=1}^n$ and all the intervals $\{(a_j,b_j)\}_{j=1}^m$ are real. However, due to the soft probability terms $\mathrm{Ps}(X=x_i)$ the result of the LHS of \eqref{soft_Expectation_def} is a soft number.
For simplicity we denote the real part of the soft expectation by $\kappa_X$, and the soft part by $\nu_X$. With this definition, the soft part $\nu_X$ adds some new information regarding to the mean of the continuous random variable $X$ given being within discrete points $\{x_i\}_{i=1}^n$. This value had been collapsed to zero without this soft expectation definition.

We can define a soft expectation of a function $g(X)$, that maps from the real numbers to the real or soft numbers as follows 
\begin{equation}\label{soft_Expectation_func_def}
\begin{split}
&\mathrm{Es}(g(X)|X\in \{x_i\}_{i=1}^n \cup \{(a_j,b_j)\}_{j=1}^m )\\ &=\sum_{i=1}^ng(x_i)\mathrm{Ps}(X=x_i) \dot{+} \sum_{j=1}^m \int_{a_j}^{b_j} g(x)f_X(x)dx\\ 
 &= \sum_{i=1}^ng(x_i)f_X(x_i)\cdot\bar{0} \dot{+} \sum_{j=1}^m\int_{a_j}^{b_j}g(x)f_X(x)dx\\
\end{split}
\end{equation}
With this concept, we define the soft variance  (denoted by $\textrm{Vs}$), related to $X$ conditionally being within union of $\{x_i\}_{i=1}^n$ and $\{(a_j,b_j)\}_{j=1}^m$. Using the nullity of $\bar{0}$ (\textbf{Axiom \ref{SN Nullity}}) and differentiation property  \eqref{softNumFunc}, we have

\begin{equation}\label{soft_Variance_def1}
\begin{split}
&\textrm{Vs}(X|X\in \{x_i\}_{i=1}^n \cup \{(a_j,b_j)\}_{j=1}^m )=\\
&\mathrm{Es}((X-(\nu_X\bar{0}  \dot{+} \kappa_X)^2)|X\in \{x_i\}_{i=1}^n \cup \{(a_j,b_j)\}_{j=1}^m )\\
&\sum_{i=1}^n[\nu_X\bar{0}
\dot{+} (\kappa_X-x_i)]^2f_X(x_i)\cdot\bar{0}\\
&\dot{+} \sum_{j=1}^m\int_{a_j}^{b_j}[\nu_x\bar{0}\dot{+} (\kappa_X-x)]^2f_X(x)dx=\\
&\left[ \sum_{i=1}^n(\kappa_X-x_i)^2 f_X(x_i) + 2\nu_X \sum_{j=1}^m\int_{a_j}^{b_j}(\kappa_X-x) f_X(x)dx \right] \bar{0}\\
&\dot{+} \sum_{j=1}^m\int_{a_j}^{b_j}(\kappa_X-x)^2 f_X(x)dx .
\end{split}
\end{equation}
We would like to simplify last equation, especially the soft part. Denote
\[\gamma_{1_X}^2=\sum_{i=1}^n(\kappa_X-x_i)^2 f_X(x_i) \geq 0,\]
\[\gamma_{2_X}=\sum_{j=1}^m\int_{a_j}^{b_j}(\kappa_X-x) f_X(x)dx \] and
\[ \lambda_X^2=\sum_{j=1}^m\int_{a_j}^{b_j}(\kappa_X-x)^2 f_X(x)dx \geq 0.\]
By the linearity of the integral, we can simplify $s_{2_X}$ as follows:
\begin{align*} 
\gamma_{2_X}&=\kappa_X\sum_{j=1}^m\int_{a_j}^{b_j}f_X(x)dx-\sum_{j=1}^m\int_{a_j}^{b_j}xf_X(x)dx \\ 
&=\kappa_X\sum_{j=1}^m[F_X(b_j)-F_X(a_j)]-\kappa_X \\
&=-\kappa_X \left[ 1 - \sum_{j=1}^m[F_X(b_j)-F_X(a_j)] \right]
\end{align*}
Observe that $1 - \sum_{j=1}^m[F_X(b_j)-F_X(a_j) > 0$.
Now we can simplify the definition for soft variance in  \eqref{soft_Variance_def1} as follows
\begin{equation}\label{soft_Variance_def2}
\begin{split}
&\textrm{Vs} (X|X\in \{x_i\}_{i=1}^n \cup \{(a_j,b_j)\}_{j=1}^m )
 =\Bigg[ \sum_{i=1}^n(\kappa_X-x_i)^2 f_X(x_i) \\
 &- 2\nu_X\kappa_X \left\{ 1 - \sum_{j=1}^m[F_X(b_j)-F_X(a_j)] \right\} \Bigg] \bar{0} \\ 
 &{\dot{+}} \sum_{j=1}^m\int_{a_j}^{b_j}(\kappa_X-x)^2 f_X(x)dx\\ 
 &=[\gamma^2_{1_X}-2\nu_X \gamma_{2_X}]\bar{0} \dot{+} \lambda^2_X\\
 &=\gamma_X\bar{0} \dot{+}\lambda^2_X.
\end{split}
\end{equation}
The real part $\lambda^2_X$ is non-negative (equals zero iif $x \equiv \kappa_X, \forall x\in(a_j,b_j),j=1,2...,m$ e.g., $X$ is deterministic), which makes sense in terms of the original definition for variance. However, in the soft part $\gamma_X=\gamma^2_{1_X}-2\nu_X \gamma_{2_X}$ we some interesting phenomena: On one hand, we have a non-negative term $\gamma^2_{1_X}$ (equals zero iif $x_i\equiv \kappa_X, \forall i=1,2...n$). On the other hand the sign of the term $-2\nu_X\gamma_{2_X}$ in the linear combination of $s_X$ depends on the sign of $\nu_X$ and the sign of $\gamma_{2_X}$ (that depends on the sign of $\kappa_X$), so that potentially the soft part may have a negative sign. Eventually, the soft part adds more information regarding to the variance of the random variable. Applications of soft variance's with negative soft max is required to be checked.
In the next subsection, we continue to define a soft entropy, inspired by the notions in this subsections.

\subsection{Soft Entropy}
Recall for the definition of the Entropy of a discrete random variable $X$ with support $S_X$ and a point mass function (PMF) $p_X$ (see e.g., \cite{Cover}) is defined by
\begin{equation}\label{Entropy_DRV_def}
\mathrm{H}(X)=-\mathrm{E}(\log p_X(X))
=-\sum_{x\in S_X} p_X(x)\log p_X(x).
\end{equation}
For a continuous case (usually referred as \textit{differential entropy}) for a continuous random variable $X\sim f_X$
\begin{equation}\label{Entropy_CRV_def}
\mathrm{H}(X)=-\mathrm{E}(\log f_X(X))
=-\int_{S_X}f_X(x)\log f_X(x)dx,
\end{equation}
where (in both definitions) the base of the logarithm operator can be chosen to be appropriate positive real number e.g. 2, $e$, 10 etc. depends on the application. In this work, we do not emphasize a specific base, but we consider like previously a continuous random variable $X$ conditioned by being within the union of $\{x_i\}_{i=1}^n$ and $\{(a_j,b_j)\}_{j=1}^m$. With these definitions, we have the following definition for a soft entropy $\mathrm{Hs}(\cdot)$:
\begin{equation}\label{soft_Entropy_def}
\begin{split}
\mathrm{Hs} & (X|X\in \{x_i\}_{i=1}^n \cup \{(a_j,b_j)\}_{j=1}^m )\\ 
 = & \sum_{i=1}^n-\mathrm{Ps}(X=x_i)\log(\mathrm{Ps}(X=x_i))\\ 
 \dot{+} &\sum_{j=1}^m \int_{a_j}^{b_j} -f_X(x)\log(f_X(x))dx\\ 
 = & \sum_{i=1}^n[-f_X(x_i)\cdot\bar{0}][\log(f_X(x_i)\cdot\bar{0})]\\
 \dot{+} &\sum_{j=1}^m\int_{a_j}^{b_j}-f_X(x)\log(f_X(x))dx\\
 = & \left [ \sum_{i=1}^n-f_X(x_i)) \right ] \cdot [\bar{0}\log(\bar{0})]  \\
 \dot{+} &  \left [ \sum_{i=1}^n-f_X(x_i)\log(f_X(x_i)) \right ] \cdot\bar{0}\\ 
 \dot{+} & \sum_{j=1}^m\int_{a_j}^{b_j}-f_X(x)\log(f_X(x))dx\\
 = & h_1\cdot [\bar{0}\log(\bar{0})] \dot{+} h_2\cdot \bar{0} \dot{+} h_3 \cdot 1, \\
\end{split}
\end{equation}
where
\[h_2\cdot \bar{0} \dot{+} h_3 \cdot 1=-\mathrm{Es}(\log(f_X(X)|X\in \{x_i\}_{i=1}^n \cup \{(a_j,b_j)\}_{j=1}^m )\]
by \eqref{soft_Expectation_func_def} with $g(x)=\log(f_X(x))$ and $h_1=\sum_{i=1}^n-f_X(x_i))$.  
The soft entropy of $X$ is a linear combination of the objects of $\bar{0}$, $\bar{0}\log(\bar{0})$ and 1. The question is how to evaluate the object $\bar{0}\log(\bar{0})$. One option would be an absolute zero i.e., $\bar{0}\log(\bar{0})$=$0$\\
\\ \textbf{Observation:}
\[\lim_{x\rightarrow 0^+}x^x=1\Rightarrow \lim_{x\rightarrow 0^+}x\log(x)=0\]
\[e^x=\sum_{n=0}^\infty\frac{x^n}{n!}=\frac{x^0}{0!}+\frac{x^1}{1!}+\frac{x^2}{2!}+\frac{x^3}{3!}+...\]
\[1=e^0=\sum_{n=0}^\infty\frac{0^n}{n!}=\frac{0^0}{0!}+(\frac{0^1}{1!}+\frac{0^2}{2!}+...)=\frac{0^0}{0!}+(0)=\frac{0^0}{1}=0^0.\]
Another option would be defined it as a new type of a "soft zero" object e.g., a new axis that is a continuum of multiples of $\bar{\bar{0}}=\bar{0}\log(\bar{0})$, with nullity rule $\bar{\bar{0}}^2=0$. With the second option, we have additional information on the Entropy of $X$, not only via the $\bar{0}$-axis but also via a new potential axis, $\bar{\bar{0}}=\bar{0}\log(\bar{0})$. 

Similarly, we define the \textit{soft cross entropy}, by evaluation of the soft expectation of $\log(\hat{f}_X(X))$ for some "guested" PDF (e.g. we assume incorrectly that $X\sim \hat{f}_X$), based on the notion of the cross entropy $\mathrm{H}(f_X,\hat{f}_X)=-\mathrm{E}(\log(\hat{f}_X(X))$ we are familiar with from Information Theory,by the following  
\begin{equation}\label{soft_cross_Entropy_def}
\begin{split}
\mathrm{Hs} & (f_X,\hat{f}_X|X\in \{x_i\}_{i=1}^n \cup \{(a_j,b_j)\}_{j=1}^m )\\ 
 = & \sum_{i=1}^n-\mathrm{Ps}(X=x_i)\log(\mathrm{\hat{P}s}(X=x_i))\\ 
 \dot{+} &\sum_{j=1}^m \int_{a_j}^{b_j} -f_X(x)\log(\hat{f}_X(x))dx\\ 
 = & \sum_{i=1}^n[-f_X(x_i)\cdot\bar{0}][\log(f_X(x_i)\cdot\bar{0})]\\
 \dot{+} &\sum_{j=1}^m\int_{a_j}^{b_j}-f_X(x)\log(\hat{f}_X(x))dx\\
 = & \left [ \sum_{i=1}^n-f_X(x_i)) \right ] \cdot [\bar{0}\log(\bar{0})]  \\
 \dot{+} &  \left [ \sum_{i=1}^n-f_X(x_i)\log(\hat{f}_X(x_i)) \right ] \cdot\bar{0}\\ 
 \dot{+} & \sum_{j=1}^m\int_{a_j}^{b_j}-f_X(x)\log(\hat{f}_X(x))dx\\
 = & \hat{h}_1\cdot [\bar{0}\log(\bar{0})] \dot{+} \hat{h}_2\cdot \bar{0} \dot{+} \hat{h}_3 \cdot 1, \\
\end{split}
\end{equation}
where
\[\hat{h}_2\cdot \bar{0} \dot{+} \hat{h}_3 \cdot 1=-\mathrm{Es}(\log(\hat{f}_X(X)|X\in \{x_i\}_{i=1}^n \cup \{(a_j,b_j)\}_{j=1}^m )\]
by \eqref{soft_Expectation_func_def} with $g(x)=\log(\hat{h}_X(x))$, $\hat{h}_1=\sum_{i=1}^n-f_X(x_i))$ and $\mathrm{\hat{P}s}(X=x)=\hat{f}_X(X)\cdot \bar{0}$ denotes the "guess" for the soft probability $\mathrm{Ps}(X=x)=f_X(X)\cdot \bar{0}$.
We can notice that the term which multiplies the object $\bar{0}\log(\bar{0})$ does not depend on the guested PDF $\hat{f}_X$. Moreover, this object is identical to the coefficient of $\bar{0}\log(\bar{0})$ in the soft entropy definition \eqref{soft_Entropy_def} (i.e., $\hat{h_1}=h_1=\sum_{i=1}^n-f_X(x_i)$). The \textit{Kullback–Leibler Divergence} (KLD, see \cite{KL} and \cite{Yu}) is defined by $\mathrm{D}(f_X||\hat{f}_X)=\mathrm{H}(f_X,\hat{f}_X)-\mathrm{H}(X)=\mathrm{E}(\log\frac{f_X(X)}{\hat{f}_X(X)})$. By subtracting  \eqref{soft_Entropy_def} from  \eqref{soft_cross_Entropy_def}, we define the \textit{soft KLD} $\mathrm{Ds}(\cdot)$ by the following
\begin{equation}\label{soft_KLD_def}
\begin{split}
\mathrm{Ds} & (f_X || \hat{f}_X|X\in \{x_i\}_{i=1}^n \cup \{(a_j,b_j)\}_{j=1}^m )\\ 
 = & \left [ \sum_{i=1}^n f_X(x_i)\log\frac{f_X(x_i)}{\hat{f}_X(x_i)} \right ] \cdot\bar{0} \\
 \dot{+} & \sum_{j=1}^m\int_{a_j}^{b_j} f_X(x) \log\frac{f_X(x)}{\hat{f}_X(x)}dx\\
 =&\mathrm{Es}\left(\log\frac{f_X(X)}{\hat{f}_X(X)}|X\in \{x_i\}_{i=1}^n \cup \{(a_j,b_j)\}_{j=1}^m \right)\\
\end{split}
\end{equation}
that has no multiple of $\bar{0}\log(\bar{0})$ term. We can see easily that that the multiple of $\bar{\bar{0}}=\bar{0}\log(\bar{0})$ term is canceled out via subtracting \eqref{soft_Entropy_def} from \eqref{soft_cross_Entropy_def}. Another explanation for it is by observing that expectation of $\log\frac{f_X(X)}{\hat{f}_X(X)}$ consists of terms with the form $\log\frac{\mathrm{Ps}(X=x_i)}{\mathrm{\hat{P}s}(X=x_i)}$. It is convenient to the perform the following cancellation of $\bar{0}$ object:
\[\log\frac{\mathrm{Ps}(X=x_i)}{\mathrm{\hat{P}s}(X=x_i)} = \log\frac{f_X(x_i)\cdot\bar{0}}{\hat{f}_X(x_i)\cdot\bar{0}}=
\log\frac{f_X(x_i)}{\hat{f}_X(x_i)}, \]
so that the multiple of $\bar{0}\log(\bar{0})$ term vanishes.

In the next section, we define a soft Mutual Information, as a splitting criteria for decision trees.

\section{Decision Trees Based on Soft Mutual Information}\label{sec5}
Decision trees (e.g. Id3, C4.5, J48 etc.) are simple yet successful techniques for predicting and explaining the relationship between some measurements about an item and its target value (see e.g., \cite{Maimon_Rokach1} and \cite{Maimon_Rokach2}). In most decision trees inducers, discrete splitting functions (also known as \textit{Splitting Criteria}) are univariate,
i.e. an internal node is split according to the value of a single attribute. Consequently, the inducer searches for the best attribute upon which to perform the split. A \textit{Splitting Criteria} of a random variable $X$ (represents the features) and a random variable $Y$ (represents the labels) has the following structure:
\begin{equation}\label{SplittingCriteria_def}
    SplittingCriteria(Y;X)=C(Y)-C(Y|X),
\end{equation}
where $C(\cdot)$ is an expectation of some cost function. In the case when the cost function is the entropy [i.e.  $C(\cdot)=\mathrm{H}(\cdot)$], we refer the splitting criteria as an \textit{Information Gain}, that is a \textit{Mutual Information} between $X$ and $Y$, denoted by
\begin{equation}\label{Mutual_Information_def}
\begin{split}
    \mathrm{I}(Y;X)&=\mathrm{H}(Y)-\mathrm{H}(Y|X)\\
          &=\mathrm{H}(Y)+\mathrm{H}(X)-\mathrm{H}(Y,X),
\end{split}    
\end{equation}
which also can be written as a KLD between the joint PDF $f_{X,Y}$ and the PDF product $f_Xf_Y$ i.e., 
\begin{equation}\label{Mutual_Information_KLD}
    \mathrm{I}(Y;X)=\mathrm{D}(f_{X,Y}||f_X f_Y).
\end{equation}

In this section we present the mutual information, an example of a splitting criteria, as a soft number,based on a joint PDF (and its related marginal PDFs) of two random continuous variable, but with a data set consist of single values and interval. From here the decision algorithm is clear. 
Suppose that $X$ and $Y$ are continuous random variables, such that $X$ is within the union of $\{x_i\}_{i=1}^n$ and $\{(a_j,b_j)\}_{j=1}^m$, and $Y$ is within the union of $\{y_i\}_{i=1}^N$ and $\{(A_j,B_j)\}_{j=1}^M$ (recall that the singles point e.g. $\{x_i\}_{i=1}^n$ and the intervals $\{(a_j,b_j)\}_{j=1}^m$ are disjoint). for simplicity denote the following sets:
\begin{equation}\label{alphabet_set}
\begin{split}
\mathcal{X}&=\{x_i\}_{i=1}^n \cup \{(a_i,b_i)\}_{i=1}^m \\
\mathcal{Y}&=\{y_j\}_{j=1}^N \cup \{(A_j,B_j)\}_{j=1}^M.
\end{split}
\end{equation}
Using \eqref{Ps(X=x, Y=y)}, \eqref{soft_KLD_def}, \eqref{Mutual_Information_KLD} and \eqref{alphabet_set}, we can define a \textit{soft Mutual Information} $\mathrm{Is}(\cdot)$ by the following equation after re-indexing

\begin{equation}\label{soft_Mutual_info_def}
\begin{split}
&\mathrm{Is}(Y;X|Y\in \mathcal{Y},X\in \mathcal{X})=\\
&\mathrm{Ds}(f_{X,Y}||f_X f_Y| Y\in \mathcal{Y},X\in \mathcal{X})=\\ 
&\left [ \sum_{j=1}^N\sum_{i=1}^n f_{X,Y}(x_i,y_j) \log\left ( \frac{f_{X,Y}(x_i,y_j)}{f_X(x_i)f_Y(y_j)}\right) \right ] \cdot\bar{0} \\
&\dot{+}\sum_{j=1}^M\sum_{i=1}^m\int_{A_j}^{B_j}\int_{a_i}^{b_i} f_{X,Y}(x,y) \log\left (\frac{f_{X,Y}(x,y)}{f_X(x)f_Y(y)}\right)dxdy, \\
\end{split}
\end{equation}
So we have an example for a splitting criteria as a soft number, that can be used in decision trees algorithms in a case of data set consist of singles values and interval.

The definition for a soft Mutual Information in \eqref{soft_Mutual_info_def}  is symmetric in $X$ and $Y$ [due to $\mathrm{I}(X;Y)=\mathrm{I}(Y;X)$ in the regular sense]. We can present \eqref{soft_Mutual_info_def} in a less symmetric form, using the Bayes Law identity $f_{X,Y}=f_{Y|X}f_X$, so that we have  

\begin{equation}\label{soft_Mutual_info_def_f_Y|X}
\begin{split}
&\mathrm{Is}(Y;X|Y\in \mathcal{Y},X\in \mathcal{X})=\\
&\left [ \sum_{j=1}^N\sum_{i=1}^n  f_{Y|X}(y_j|x_i)f_X(x_i) \log\left ( \frac{f_{Y|X}(y_j|x_i)}{f_Y(y_j)}\right) \right ] \cdot\bar{0}\\
&\dot{+} \sum_{j=1}^M\sum_{i=1}^m\int_{A_j}^{B_j}\int_{a_i}^{b_i} f_{Y|X}(y|x)f_X(x) \log\left (\frac{f_{Y|X}(y|x)}{f_Y(y)}\right)dxdy.
\end{split}
\end{equation}    
This representation is applicable e.g., for emphasizing $X$ as an input and $Y$ as an output to some channel. An example is shown in the next subsection for a Gaussian case.

\subsection{Gaussian Distribution Example}
Consider the jointly Gaussian distributed variables $X$ and $Y$ as follows:
\begin{equation}\label{Gaussian variables X & Y}
\begin{split}
X&\sim N(0,1), f_X(x)=\frac{1}{\sqrt{2\pi}} e^{-\frac{1}{2}x^2}\\
Y&\sim N(0,2), f_Y(y)=\frac{1}{\sqrt{2\pi \cdot 2}} e^{-\frac{1}{2\cdot 2}y^2}\\
(Y|X=x)&\sim N(x,1),f_{Y|X}(y|x)=\frac{1}{\sqrt{2\pi}} e^{-\frac{1}{2}(y-x)^2}\\
\end{split}
\end{equation}

\begin{remark}\label{remark1}
The above setup can be obtained by adding an uncorrelated Gaussian noise $W\sim N(0,1)$ to the Gaussian input $X$ such that $X \indep W$ and we have $Y=X+W$. A sketch of the proof is shown below 
\begin{align*}
    \mathrm{E}(Y)&=\mathrm{E}(X)+\mathrm{E}(W)\\&=0+0\\&=0,
\end{align*}
\begin{align*}
    \mathrm{Var}(Y)&\overset{X \indep W}{=}\mathrm{Var}(X)+\mathrm{Var}(W)\\ &=1+1 \\&=2,
\end{align*}
\begin{align*}
    \mathrm{E}(Y|X=x)&=\mathrm{E}(X|X=x)+\mathrm{E}(W|=x)\\&\overset{X \indep W}{=}x+0\\&=x,
\end{align*}
\begin{align*}
    \mathrm{Var}(Y|X=x)&=\mathrm{Var}(Y-x|X=x)\\&=\mathrm{Var}(W|X=x)\\&\overset{X \indep W}{=}1.
\end{align*}
                     
We used Gaussian distributions in this example due to the properties of jointly Gaussian random variables. We can use any continuous distributions and to calculate the soft mutual information accordingly. 
\end{remark}

Consider a simple case that each set $\mathcal{X}$ (input set) $\mathcal{Y}$ (output set) have one open interval and one single point
\begin{equation}\label{X & Y input and output set}
\begin{split}
\mathcal{X}&=(a,b)\cup \{x_0\},\\
\mathcal{Y}&=(A,B)\cup \{y_0\},\\
\end{split}
\end{equation}
so that the soft Mutual information in  \eqref{soft_Mutual_info_def_f_Y|X} is given by
\begin{equation}\label{soft_Mutual_info_one_pt_intrvl}
\begin{split}
&\mathrm{Is}(Y;X|Y\in \mathcal{Y},X\in \mathcal{X})=\\
&\left [  f_{Y|X}(y_0|x_0)f_X(x_0) \log\left ( \frac{f_{Y|X}(y_0|x_0)}{f_Y(y_0)}\right) \right ] \cdot\bar{0} \\
&\dot{+}  \int_{A}^{B}\int_{a}^{b} f_{Y|X}(y|x)f_X(x) \log\left (\frac{f_{Y|X}(y|x)}{f_Y(y)}\right)dxdy. \\
\end{split}
\end{equation} 
and after plugging the Gaussian PDFs [according to \eqref{Gaussian variables X & Y}], we have
\begin{equation}\label{soft_Mutual_info_Gauss}
\begin{split}
&\mathrm{Is}(Y;X|Y\in \mathcal{Y},X\in \mathcal{X})=\\
&\left [  \frac{1}{2\pi} e^{-\frac{1}{2}(y_0-x_0)^2}e^{-\frac{1}{2}x_0^2} \log\left ( \sqrt{2} e^{\frac{1}{2\cdot 2}y_0^2} e^{-\frac{1}{2}(y_0-x_0)^2}   \right) \right ] \cdot\bar{0} \\
 &\dot{+}  \int_{A}^{B}\int_{a}^{b} \frac{1}{2\pi} e^{-\frac{1}{2}(y-x)^2}e^{-\frac{1}{2}x^2} \log\left ( \sqrt{2} e^{\frac{1}{2\cdot 2}y^2} e^{-\frac{1}{2}(y-x)^2}   \right)dxdy.
\end{split}
\end{equation}

At the following Table \ref{table:1}, we obtain some numerical results for a Soft Mutual Information (denoted by $\mathrm{Is}(Y;X)$ for simplicity) in our Gaussian Case. We used logarithm with base $e$: 
\begin{table}[H]
\caption{Numerical results of a Soft Mutual Information in the Gaussian Case}
\centering
\begin{tabular}{||c c c c c||} 
 \hline
 $x_0$ & $y_0$ & $(a,b)$ & $(A,B)$ & $\mathrm{Is}(Y;X)$ \\ [0.5ex] 
 \hline
 0 & 0 & (1,2) & (1,2) & 0.055159$\cdot\bar{0}$ $\dot{+}$ 0.042381 \\ 
 0 & 1 & (1,2) & (2,3) & 0.0093225$\cdot\bar{0}$ $\dot{+}$ 0.037941 \\ 
 1 & 0 & (2,3) & (1,3) & -0.0089831$\cdot\bar{0}$ $\dot{+}$ 0.018353 \\ 
 1 & 0 & (20,30) & (10,30) & -0.0089831$\cdot\bar{0}$ $\dot{+}$ 2.7404E-87 \\ 
 20 & 30 & (2,3) & (1,3) & 7.4494E-108$\cdot\bar{0}$ $\dot{+}$ 0.018353 \\ 
 \hline
\end{tabular}
\label{table:1}
\end{table}
We can observe that, on one hand, when $x_0$ and $y_0$ are far away from the mean of $X$ and $Y$ (zero for both in our case), the contribution of the soft mutual information in its soft part approaches to zero. On the other hand, when the intervals $(a,b)$ and $(A,B)$ are away from the mean of $X$ and $Y$, the contribution of the soft mutual information in its real part approaches to zero, so the soft part may have a significant value for taking a decision in a soft decision tree.

To summarize this example, we consider a case of two jointly Gaussian variables. We generate a formula for a Soft Mutual Information in a simple case of when each random variable's datum consists of one single point (to generate the soft part of the Soft Mutual Information) and one interval (to generate the real part of the Soft Mutual Information). This example can be generalized by summing the contributions of the Soft Mutual Information of any set of disjoints singles points and intervals.

\section{Conclusions}\label{sec6}
In the classical probability, in continuous random variables there is no distinguishing between the probability involving strict inequality and non-strict inequality. Moreover, a probability involve equality collapse to zero, without distinguishing among the values that we would like that the random variable will have for comparison. Soft numbers assist us to distinguish between the probability involving strict inequality and non-strict inequality, and among the values that we would like that the random variable, by generating soft zeros multiples of the PDF observations.

In addition, we extended this notion of soft probabilities to the classical definitions of Complements, Unions, Intersections and Conditional probabilities under Kolmogorov definition and Bayes theorem, that makes sense with a probability of a continuous variable to be equal to an exact value does not collapse completely to zero.

We also extend the notion of soft probabilities to the expectation, variance and entropy of a continuous random variable, condition being in a union of disjoint intervals and a discrete set of numbers. with this extension, we have some information regarding to the expectation, variance and entropy of a continuous random variable being within discrete sent of numbers, but not collapse completely to zero. In addition we discover some interesting properties regarding to soft variance and soft entropy that required to be explored. In soft variance, the soft part might be a negative number. In the soft entropy, we have potentially a new zero axis with multiples of $\bar{0}\log(\bar{0})$, or alternatively we may defined $\bar{0}\log(\bar{0})$ as an absolute zero. For the first option (considering new zero axis) it may be required to define additional bridging notation in order to bridge between multiples of $\bar{0}\log(\bar{0})$
and the multiples of 0 and 1. 
We extended the notion of soft entropy into the definition of Cross Entropy and KLD, and we found that a soft KLD is a soft number, that does not have a multiple of $0\cdot\log0$. More exploration are required to be done in order to realize the consequences  of this result. Based on a soft KLD, we defined a soft mutual information, that can be used as a splitting criteria in decision trees with data set of continuous random variables, consist of single samples and intervals.

 \section{Suggestions for Future Research}\label{sec7}
 We suggest to extend the notion of soft probability covered in this work by generalizing to the followings: continuous random vectors, mixed random variable (that has continuous and discrete distribution i.e., non piecewise constant CDF but with discontinuity), random vector with discrete, continuous and mixed random variables etc.
 
 In addition we suggest to explore the applications of negative soft part in the soft variances, and to explore the applications of multiples of $\bar{0}\log(\bar{0})$ as an information to the soft entropy (in addition to the information regarding to the multiples of multiples of $\bar{0}$ in the soft entropy). We also suggest to explore the soft logic in general and soft probability in particular in additional topics in information theory, data mining, machine learning, computability, meta-verse technology, cyber-physical system (CPS) etc. We also suggest to involve the views of the theory of consciousness in the mentioned above scientific and technological topics, with the concept of the zero axis presents the inner world or virtual world, and the one axis the real world (see paragraph below \eqref{SP_sol} for more details).
 We believe that with soft logic (and soft probability) we can incorporate the spiritual concept of consciousness, that present inner/virtual world or the zero axis, into the scientific and technological topics in the real world or the one axis. 

\appendices

\renewcommand{\thesection}{\Alph{section}}
\numberwithin{equation}{section}

  \section{Probability Theory Brief Review}
  \label{Appendix A}
Probability theory is used in order to model processes and phenomenons, involving randomness of the parameters and variables. Usually, when we want to quantify a probability of an event in these processes or phenomenons, we evaluate the probability of this event by the range [0,1], e.g., '0' means the event can never (almost surely) occur and 1 means the event can always (almost surely) occur. For this quantification, a probability space is a defined by mathematical triplet  \((\Omega, \mathcal{F}, P)\) defined as follows:
\begin{itemize}
\item Sample Space $\Omega$: Set of all possible outcomes. An outcome is the result of a single execution of the model.
\item $\sigma$-algebra $\mathcal{F}$: Collection of all the events we would like to consider. An event is a set outcomes.  
\item Probability Measure $P$: Function returning an event's probability. $P$ maps from the $\sigma$-algebra $\mathcal{F}$ to the interval [0,1].
\end{itemize}
Random variables are used to provide outcomes numerical values. The mathematical notation for a random variable $X$ is defined by the following:
\begin{equation}
X:\Omega\rightarrow S_X 
\end{equation}
where $S_X$ is a set of real numbers, that the random variable $S_X$ can have. $S_X$ is called the \textit{support of} $X$. 
Mainly, we distinct between two types of Random variables:
\begin{itemize}
\item Discrete random variables, that can have finite or countable of values; and
\item Continuous random variables that can have uncountable of values. 
\end{itemize}
For both discrete and continuous random variables, a cumulative distribution function (CDF) of a random variable $X$ as follows:
\begin{equation}\label{CDF}
F_X(x)=\mathrm{Pr}(X\leq x), 
\end{equation}
where the right hand side (RHS) asked what is the probability of a random variable $X$ to be less or equal to some real number $x$, and the left hand side (LHS) provides the answer in terms of $x$ by the function $F_X:\mathbb{R} \rightarrow[0,1]$.   

In a case of a discrete random variable, we can address to the question, what is the probability of a random variable $X$ to be equal to some real number $x$, by the point mass function (PMF), defined as follows: 
\begin{equation}
p_X(x)=\mathrm{Pr}(X=x). 
\end{equation}
The cumulative property is obtained by the following relation between the CDF and the PMF in the discrete case
\begin{equation}
F_X(b)-F_X(a)=\sum_{a<x_i\leq b}p_X(x_i)=\mathrm{Pr}(a<X\leq b). 
\end{equation}

In a case of a continuous random variable, a probability density function (PDF) is defined by:
\begin{equation}
f_X(x)=\frac{dF_X(x)}{dx}. 
\end{equation}
The function $f_X:\mathbb{R}\rightarrow \mathbb{R}_{\geq 0}$ denotes the PDF of $X$. The cumulative property is obtained by the following relation between the CDF and the PDF in the continuous case
\begin{equation}
F_X(b)-F_X(a)=\int_{a}^{b}f_X(x)dx=\mathrm{Pr}(a<X\leq b). 
\end{equation}
The PDF can be used is to approximate the probability of the continuous random variable $X$ to be adjacent to $x$ in the following sense
\begin{equation}\label{Prob_approxs}
\mathrm{Pr}(x<X\leq x+\Delta x) \approx f_X(x)\Delta x,  
\end{equation}
where $\Delta x>0$ is  a small value, that defines how much this probability is accurate. However, continuous random variables have the following properties:
\begin{itemize}
\item No distinguishing between strict inequality and non-strict in equality e.g., $\mathrm{Pr}(X\leq x)=\mathrm{Pr}(X<x)$;
\item Equality collapses to zero i.e.,  $\mathrm{Pr}(X=x)=0$. Although any value of $x\in S_X$ is possible for $X$, the the probability of $X$ to be equal to any value of $x\in S_X$ is (almost surely) zero.  
\end{itemize}
Because of these properties, we lose some information regarding to a continuous random variable to have an exact value. 

In the literature (e.g., \cite{Li_Chen}, \cite{Khuri} and \cite{Zanetti_Tuggle}), there is an approach to represent a discrete distribution as a continuous distribution by a linear combination of Dirac delta functions $\delta(x-x_i)$, or by any approximations of Dirac delta functions e.g., Gaussian functions (also known as Gaussian mixture model or GMM) or rectangular functions (based on uniform distribution) etc. Suppose $X$ is a discrete random variable with the probability $\mathrm{Pr}(X=x_i)=p_i$. Then $X$ can be represented with a continuous distribution as follows

\begin{equation}\label{discrete_t_delta}
\begin{split}
    f_X(x)& =\sum_i p_i\delta(x-x_i)\\
    &\approx\sum_i p_i\cdot \frac{1}{\sqrt{2\pi\sigma^2}}e^{-\frac{1}{2\sigma^2}(x-x_i)^2},\sigma^2 \ll 1\\
    &\approx\sum_i p_i \cdot\frac{1}{2a}\mathbbm{1}_{x-x_i\in (-a,a)},a \ll 1.\\
\end{split}    
\end{equation}
recall that, 
\begin{equation}
\begin{split}
&\delta(x)=\begin{cases} 
      0 & x\neq 0 \\
      \infty & x=0 \\
  \end{cases}
  \\
&\int_{-\infty}^{\infty} \delta(x)dx =1,
\end{split}  
\end{equation}
and also $\frac{1}{\sqrt{2\pi\sigma^2}}e^{-\frac{1}{2\sigma^2}x^2}\xrightarrow[]{\sigma^2\rightarrow 0}\delta(x)$, $\frac{1}{2a}\mathbbm{1}_{x\in (-a,a)}\xrightarrow[]{a\rightarrow 0}\delta(x)$, i.e., Gaussian distribution and uniformly distribution converge to Dirac delta function (degenerative distribution) when the variance of the Gaussian distribution and the length of the interval in the uniformly distribution approach to zero respectively. Our approach it to establish the opposite in some sense, i.e., to represent a continuous random variable with a possibility to have a discrete values with probability that will not collapse absolutely to zero.

In this work, we introduce the \textit{Soft Numbers} (see Klein and Maimon's  papers e.g., \cite{Klein_Maimon1}, \cite{Klein_Maimon2} and \cite{Klein_Maimon3}) to give a probability interpretation of a continuous random variable to have an exact value, that provides distinguishing between strict inequality and non-strict in equality in the probability function.

 \section{Presentation of Soft Numbers}
  \label{Appendix B}
According to traditional mathematics, the expression 0/0 is undefined, although in fact the whole set of real numbers could represent this expression, since $a\cdot0= 0$ for all real numbers $a$. This observation opens a new area for investigation, which is a part of what it is called in \cite{Klein_Maimon1} a “Soft Logic", that refers to a new axis, "a continuum of multiples of zeros",  with distinction between a positive zero "+0" and a negative zero "-0" (see also \cite{Klein_Maimon2} and \cite{Klein_Maimon3}).  

\subsection{Soft Number: Definitions and Axioms}
A new object $\bar{0}$ is symbolized in order to generate of a continuum of multiples of zeros $a\bar{0}$ on a "$\bar{0}$" axis, where $a$ is a real number. An object $a\bar{0}$ denotes "soft zero", while the object $\textbf{0}=0\cdot \bar{0}$ denotes "absolute zero".  The object $\bar{1}$ denotes the real axis (i.e., contains multuples of "ones", $b\bar{1}$),  and parallel to the "$\bar{0}$" axis. For simplicity, the symbol $\bar{1}$ is omitted during computations.The following axioms and definitions are developed  for soft zeros for all real numbers $a$ and $b$:
\begin{axiom}[Distinction]\label{SN Distinction} 
$a\neq b\Rightarrow a\bar{0}\neq b\bar{0}$.
\end{axiom}
\begin{definition}[Order] \label{SN Order} $a< b\Rightarrow a\bar{0}< b\bar{0}$.
\end{definition}
\begin{axiom}[Addition]\label{SN Addition}
$a\bar{0}+b\bar{0}=(a+b)\bar{0}$.
\end{axiom}
\begin{axiom}[Nullity]\label{SN Nullity}
$a\bar{0}\cdot b\bar{0}=0$, i.e., soft numbers "collapse" to zero under multiplications.
\end{axiom}
\begin{axiom}[Bridging] \label{Bridge No.}
There exists a bridge between a zero axis, and a real axis and vice versa, denoted by a pair of a bridge number and its mirror image about the bridge sign. Bridge numbers of a right type
\[
b\bar{1}\perp a\bar{0}
\]
and  bridge numbers of a left type 
\[
a\bar{0} \perp b\bar{1}.
\]
\end{axiom}
\begin{axiom}[Non-commutativity]\label{Bridge Not commute}
Bridging operator $\perp$ does not commute \cite{Klein_Maimon3} i.e.,  
\[b\bar{1}\perp a\bar{0} \neq a\bar{0} \perp b\bar{1}.\]
\end{axiom}
\begin{definition}[Soft Number]\label{SN def} \textit{A soft number is defined as a set of the of bridge numbers pair of opposite types but with the same components – the same zero axis number $a\bar{0}$ and the same real number $b$:}
\[
a\bar{0} \dot{+}b=\{a\bar{0}\perp b;b\perp a\bar{0}\}
\]
\end{definition}

\begin{figure}[ht]
    \centering
    \includegraphics{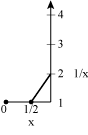}
    \caption{The Soft coordinate axis}
    \label{fig1}
\end{figure}

We denote the set of all bridge numbers by $\textbf{BN}$ and all soft numbers by $\textbf{SN}$. The coordinate system of Soft Logic is constructed, as presented in Figure \ref{fig1}. It starts from 0 to 1 horizontally and then it turns 90$^\circ$ from 1 to infinity  

\begin{remark}\label{remark2}
There exists a one-to-one correspondence between the segment $(0, 1]$ and the segment $[1,\infty)$.
\end{remark}

\begin{remark}\label{remark3}
All lines that connect $x$ to $1/x$ (for all non-zero real $x$) intersect at a single point.
\end{remark}

\begin{figure}[H]
    \centering
    \includegraphics{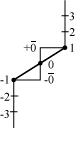}
    \caption{Distinction between -0 and +0}
    \label{fig2}
\end{figure}

The statements in Remarks \ref{remark2} and \ref{remark3} were demonstrated in \cite{Klein_Maimon1}. This “single point” denotes the beginning of the soft logic coordinate system. We call this point “the absolute zero”. The distance from absolute zero to +0 is 1. An extension of this new coordinate system to the negative numbers is implemented in Figure \ref{fig2}.

In Figure \ref{fig2} we have, in addition to the absolute zero \textbf{0}, two additional zeros. One zero is opposite the number $-1$, and is not identical with the zero opposite to the number $+1$. Hence, we suggest denoting these two different "zeros" as $+\bar{\textbf{0}}$ and $-\bar{\textbf{0}}$.

\begin{figure}[ht]
    \centering
    \includegraphics{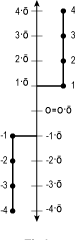}
    \caption{The extended soft coordinate system}
    \label{fig3}
\end{figure}

Figure \ref{fig3} shows the extended coordinate system for positive and negative numbers with an additional line presenting the multiples of zero. The added line is called a zero line or a zero axis, and the multiples on it are called soft zeros or zero axis numbers.

\begin{figure}[ht]
    \centering
    \includegraphics{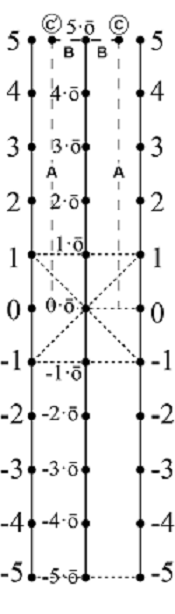}
    \caption{The complete soft coordinate system}
    \label{fig4}
\end{figure}

The coordinate system in Figure \ref{fig3} allows us to present all the real numbers and all the soft zeros. We now wish to construct a coordinate system for representing various Soft Numbers, which may be described as an infinite strip as shown in Figure 4. Because of the Soft Number duality, we double the strip (Figure \ref{fig4}). This allows us to represent both elements of a Soft Number: 
\begin{equation}
\begin{split}
c &=x\bar{0} \perp y,\\
c'&=y\perp x\bar{0},
\end{split}
\end{equation}
where $x$ and $y$ are real numbers. Each of the elements $c$ and $c'$ is a mirror image of the other about the bridge sign.
Note that we have expanded the coordinate system in Figure \ref{fig3} to the one shown in Figure \ref{fig4}.

As the infinite strip, presented (partially) in Figure \ref{fig4}, is intended for the presentation of Soft Numbers, we call it a \textit{‘Soft Numbers Strip’} or briefly, SNS.
\begin{definition}[height and width of a point on an SNS]
let $C$ be any point on the SNS.
\begin{itemize}

\item The \textbf{height of the point} $C$ is the vertical distance from $C$ to the horizontal segment with the absolute zero at its center. This distance is supplied with a plus sign if $C$ is above this segment and with a minus sign if $C$ is below it. The height with a sign is denoted by $A$.

\item The \textbf{width of the point} $C$
is the horizontal distance from $C$ to the zero line and is denoted by $B$.

\end{itemize}

\end{definition}

The definitions above provide every point C on the SNS with two parameters, $A\in \mathbb{R}$ and $B\in[0,1]$. The condition  $A>0$  is satisfied in the positive part of the SNS, and $A<0$ - in its negative part, or correspondingly, above and below the horizontal segment containing the absolute zero, while on this segment $A=0$. For the second parameter $B$ there is: $B=0$ on the zero axis, $B=1$ on the lines bounding the SNS, and otherwise $0<B<1$.

If two points $c$ and $c'$ on the SNS are symmetric about the zero axis, they have the same height $A$ and the same width $B$, i.e., we can symmetrically represent them by the following $\textbf{BN}$s:
\begin{equation}
\begin{split}
    &c =(1-B)A\bar{0}\perp BA,\\
    &c'=BA\perp (1-B)A\bar{0}.
\end{split}
\end{equation}
Therefore, to define a presentation of soft numbers $x\bar{0}\dot{+}y$ by symmetric pairs ($\textbf{SP}$s) of points on the SNS, we have to define a correspondence between these numbers and the pairs of real numbers $(A, B) \in\mathbb{R} \times [0,1]$ (denoted as $\textbf{SP}$), so that
\begin{equation}
\begin{split}
    x\bar{0}\dot{+}y &= \{c,c'\}\\
                     &=(1-B)A\bar{0} \dot{+} BA.
\end{split}
\end{equation}
Hence, by a coefficients comparison of the real part and the soft part:
\begin{equation}
\begin{split}
    x &= (1-B)A\\
    y &= BA,
\end{split}
\end{equation}
or equivalently, after solving for the $\textbf{SP}$, $(A, B)$
\begin{equation}\label{SP_sol}
\begin{split}
    A &= x+y\\
    B &= \frac{y}{x+y}.
\end{split}
\end{equation}

It can be proven that there is an algebraic isomorphism between the bridge numbers $b\bar{0}\perp a$ and Dual numbers developed by Clifford \cite{Clifford} with the form $a+b\varepsilon$, where $\varepsilon^2=0$ but $\varepsilon \neq 0$. The main difference between $\varepsilon$ in Dual numbers and $\bar{0}$ is the realisation and geometrical interpretation of $\bar{0}$  as an extension of the number 0 on a continuous line. This line can be a model of the inner world, while $\bar{1}$ is a model of the real world. The bridge between them enables us to treat the concept of consciousness with mathematical tools. Another difference is the possibility, in Soft logic, of developing a Soft curve \cite{Klein_Maimon3}.

One of our major topics for investigation in further research is the connection of soft numbers to Mobius strip. In order to describe the geometry of Mobius strip with soft numbers,we suggest to modify the soft coordinate system in Figure \ref{fig4} by alternating the sign of left vertical line.

\begin{figure}[H]
    \centering
    \includegraphics{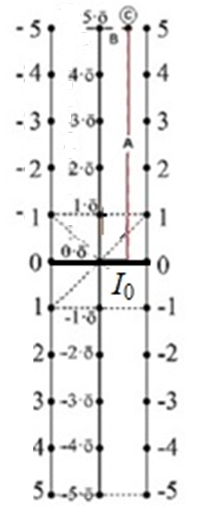}
    \caption{The alternative soft coordinate system}
    \label{fig5}
\end{figure}
The horizontal line $I_0$ in Figure \ref{fig5} can represent the connection line where the edges of a straight strip is twisted and attached together to create a Mobius strip. One of the suggestion to define a point on the Mobius strip with soft numbers is that $c =(1-B)A\bar{0}\perp BA$ is located in the front of this page, while $c'=BA\perp (1-B)A\bar{0}$ is located behind this page. This setup demonstrates locally existence of two side of Mobius strip. However, it is known that Mobius strip has globally one side. Moreover, if we start walking vertically from the  point $c$ ($A$ units from $I_0$ and $B$ units from the zero axis) on the front of this page, we will pass through the point behind this page but across the point $c'$ and ($-A$ units from $I_0$ and $B$ units from the zero axis). When we keep walking on that point, we will go back to the starting point $c'$. Because of this phenomenon, we are motivated to explore the possibility to represent a soft number with more than two symbols. 

In the next subsection, we outline some properties of mathematical operations and functions over the soft numbers. 

\subsection{Mathematical operations and Functions on Soft Numbers}
In this section we outline some mathematical operations over the soft numbers. Suppose $a \bar{0} \dot{+}b, c \bar{0} \dot{+}d \in \textbf{SN}$ are given soft numbers, then the following mathematical operations hold based on axioms \ref{SN Addition} and \ref{SN Nullity}:
\begin{itemize}
\item \textbf{Addition/subtraction:}
\begin{equation}
(a\bar{0} \dot{+}b)\pm (c\bar{0} \dot{+}d)=(a\pm c)\bar{0} \dot{+}(b\pm d);
\end{equation}
\item \textbf{Multiplication:}
\begin{equation}
(a\bar{0} \dot{+}b)\cdot (c\bar{0} \dot{+}d)=(ad+bc)\bar{0} \dot{+}bd; 
\end{equation}
\item \textbf{Natural power:}
\begin{equation}
(a\bar{0} \dot{+}b)^n=nab^{n-1}\bar{0} \dot{+}b^n.
\end{equation} 
\end{itemize}

Based on the above equations, every polynomial $P_N(x)$ that operates on every soft number $\alpha \bar{0} \dot{+}x$ is given by
\begin{equation}
P_N(\alpha \bar{0} \dot{+}x)=\alpha P_N'(x)\bar{0} \dot{+}P_N(x).
\end{equation}
where $P_N'(x)$ denotes the derivative of $P_N(x)$. This notion is generalized for analytic functions $f(x)$ so that
\begin{equation}\label{softNumFunc}
f(\alpha \bar{0} \dot{+}x)=\alpha f'(x)\bar{0} \dot{+}f(x).
\end{equation}



\section*{Acknowledgment}

This paper was supported by the Koret Foundation grant for Smart Cities and Digital Living 2030 bestowed upon the universities of Stanford and Tel Aviv. We are happy to express our thankfulness and gratitude for this support.


\end{document}